\documentclass[conference]{IEEEtran}
\IEEEoverridecommandlockouts

\usepackage{amsmath,graphicx}
\usepackage{url}
\usepackage{xcolor}
\usepackage{float}
\usepackage{soul}
\usepackage{url}
\usepackage{verbatim}
\usepackage{graphicx}
\usepackage{subcaption}
\usepackage{amsmath}
\usepackage[table]{colortbl}
\usepackage{algorithm}
\usepackage{algpseudocode}
\usepackage{cite}
\usepackage{amsmath,amssymb,amsfonts}
\usepackage{comment}
\usepackage[most]{tcolorbox}

\title{SONIC: Synergizing VisiON Foundation Models for Stress RecogNItion from ECG signals}
\author
{Orchid Chetia Phukan$^{1\ast \dagger}$, Ankita Das$^{2\ast}$,  Arun Balaji Buduru$^{1}$, Rajesh Sharma$^{1,3}$\\
\normalsize{$^{1}$IIIT-Delhi, India}\\
\normalsize{$^{2}$IIIT-Guwahati, India}\\
\normalsize{$^{3}$University of Tartu, Estonia}\\
\normalsize{$^\dagger$orchidp@iiitd.ac.in}
}

\begin{document}
%
\maketitle
\def\thefootnote{*}\footnotetext{The authors contributed equally to this work}
\begin{abstract}
Stress recognition through physiological signals such as Electrocardiogram (ECG) signals has garnered significant attention. Traditionally, research in this field predominantly focused on utilizing handcrafted features or raw signals as inputs for learning algorithms. However, there is now a burgeoning interest within the community in leveraging large-scale vision foundation models (VFMs) like ResNet50, VGG19, and others. These VFMs are increasingly preferred due to their ability to capture complex features, enhancing the accuracy and effectiveness of stress recognition systems. However, no particular focus has been given on combining these VFMs. The combination of VFMs offers promising benefits by harnessing their collective knowledge to extract richer representations for improved stress recognition. So, to mitigate this research gap, we focus on combining different VFMs for stress recognition from ECG and propose SONIC, a novel framework that combines VFMs through their logits and training a fully connected network on the combined logits. Through extensive experimentation, SONIC showed the top performance against individual VFMs performance on the WESAD benchmark. With SONIC, we report state-of-the-art (SOTA) performance in WESAD with 99.36\% and 99.24\% (stress \textit{vs} non-stress) and 97.66\% and 97.10\% (amusement \textit{vs} stress \textit{vs} baseline) in accuracy and F1 respectively.

\end{abstract}

\section{Introduction}
\label{sec:intro}

Stress is a human affective condition and it refers to the physiological and psychological response that arises when an individual perceives a potential threat, challenge, or demand in their environment. Stress is not always detrimental; rather it can serve as a mechanism to help us cope with different situations. However, prolonged chronic stress can have adverse effects on physical well-being such as increased heart rate \cite{cohen2007psychological}, increased blood pressure \cite{hahad2019environmental}, cancer \cite{mravec2020stress}, etc., and mental well-being leading to cognitive problems such as anxiety, depression \cite{hahad2019environmental}, etc. Therefore, recognizing stress at an early stage is essential to prevent it from becoming chronic. By fostering increased awareness of elevated stress levels that could otherwise go unreported, technology for stress recognition has the potential to improve people's understanding of stress and facilitate its treatment.\par

Traditionally, experts in psychology and physiology used a questionnaire-based stress analysis\cite{crosswell2020best} to assess a person's level of stress. However, this technique lacks reliability because it only uses the subject's self-reported responses. The accuracy of the assessment may also be hampered by people's hesitation when answering the questionnaire. To overcome these limitations, researchers have investigated various means such as voice \cite{wu23i_interspeech}, facial landmarks \cite{naidu2021stress}, physiological signals captured by wearable devices \cite{tanwar2024attention}, etc. In this work, we specifically concentrate on one of the commonly used physiological signals used for stress recognition i.e. ECG (Electrocardiogram). \par

Stress recognition from ECG has largely benefitted from the availability of various large-scale datasets such as WESAD \cite{schmidt2018introducing}, StressID \cite{chaptoukaev2024stressid}, and so on. Initial works involved usage of classical machine learning (ML) algorithms such as Random Forest, AdaBoost, etc. \cite{schmidt2018introducing, benchekroun2022comparison} with handcrafted features as well as deep learning techniques such as CNN \cite{giannakakis2019novel, zhang2021real, zhang2021psychological, seo2019deep}. Researchers also built a transformer-based model that comprises of a convolutional block followed by a transformer encoder achieving state-of-the-art (SOTA) performance for stress recognition from ECG \cite{behinaein2021transformer}. \par

Moreover, there is a rising interest in the community towards the utilization of large-scale vision foundation models (VFMs) such as ResNet50, VGG19, and similar architectures \cite{ishaque2022detecting} in stress recognition from ECG. These VFMs despite being trained primarily on visual data provide enhanced performance in detecting stress through ECG. Additionally, this contributes to saving both time as well as resources and prevents training models from scratch. VFMs come in different types of model architectures. They can be either CNN-based \cite{simonyan2014very} or transformer-based \cite{dosovitskiy2020image} architectures and pre-trained on very-large-scale visual data. \par

However, no attention towards combining these VFMs has been given by previous works for stress recognition from ECG. Combining VFMs with distinct architectural strengths comes with improved and robust performance also shown in tasks in medical imaging \cite{bakasa2023stacked, alotaibi2023vit}. Each VFM can excel at capturing certain aspects of input patterns due to variations in their training data and architectural design. By combining VFMs, we can capture a broader spectrum of features, enabling a more holistic representation of the input features. This rich feature representation can significantly improve the discrimination between stress and non-stress signals. Furthermore, combining VFMs can facilitate complementary behavior, wherein the strengths of individual VFMs complement each other while their weaknesses are mitigated. To cover this research gap, we focus on the combination of VFMs and propose \textbf{SONIC} (\textbf{S}ynergizing Visi\textbf{ON} Foundation Models
for Stress Recog\textbf{NI}tion from Ele\textbf{C}trocardiogram
Signals), a novel modeling framework that merges VFMs by fusing their logits and training a fully connected network (FCN) using the fused logits for improved stress recognition from ECG. 

\noindent {The main contributions are summarized as follows:}
\begin{itemize}
  \item We propose a novel framework, namely \textbf{SONIC}, which illustrates methodology to combine selected VFMs by fusing their logits. 
  \item With \textbf{SONIC}, we report SOTA performance in terms of accuracy and F1 in 2-class classification (stress \textit{vs} non-stress) as well as 3-class classification (amusement \textit{vs} stress \textit{vs} baseline). The proposed method reports almost 9\% and 20\% improvement in accuracy in comparison to previous SOTA works in 2-class and 3-class classification. 
\end{itemize}

\noindent We will release the codes and models curated as part of our study after the review process to maintain reproducibility and reference for future works to build upon our work.

\section{Vision Foundation Models}

In this section, we discuss the VFMs considered for our study. We select these VFMs as they have substantially led to improvements in ImageNet performance as well as in various downstream tasks. They are as follows:

\noindent\textbf{VGG19\cite{simonyan2014very}:} It's a part of the VGG family of models built for large-scale image recognition and was proposed as a model to test the hypothesis that increasing model depth leads to improvements in performance. VGG-19 comprises of 19 layers in total and trained on ImageNet dataset. Its architecture consists of 16 convolutional layers followed by 3 fully connected layers. Central to its design, VGG-19 employs small 3x3 convolutional filters, which are effective in capturing detailed features within images. This choice of filter size contributes to the network's ability to discern intricate patterns, enhancing its performance in image recognition tasks. \par

\noindent\textbf{ResNet50\cite{he2016deep}:} It is a CNN comprising of 50 layers and trained on ImageNet. Its distinguishing feature lies in the incorporation of residual connections or skip connections and these connections enable the network to bypass one or more layers, thereby mitigating the vanishing gradient problem commonly encountered in deep neural networks. Furthermore, skip connections play a crucial role in optimization by enabling the model to converge to a better local minimum. By providing alternative paths for gradient flow, ResNet-50 can navigate through the parameter space more efficiently, leading to improved convergence and potentially higher performance.

\noindent\textbf{EfficientNet \cite{tan2019efficientnet}:} It is groundbreaking CNN architecture trained on ImageNet and designed for superior performance and computational efficiency. Its innovation lies in the implementation of compound scaling, a method that uniformly adjusts the network's depth, width, and resolution to strike an optimal balance between model size and accuracy across diverse computational constraints. With variants ranging from EfficientNet-B0 to B7, each is meticulously scaled for different resource requirements.\par

\noindent\textbf{ViT \cite{dosovitskiy2020image}:} It is a revolutionary convolution-free model architecture built upon original transformer \cite{vaswani2017attention} architecture. However, contrary to the original transformer architecture built for NLP tasks, ViT leverages only the encoder part of it. It replaces the convolution networks with self-attention heads. It is designed primarily for image classification tasks and trained on ImageNet and showed competitive performance in comparison to its previous ConvNet architectures. ViT divides an image into smaller patches of a fixed size instead of processing an entire image at once. These patches are transformed into vectors through linear embeddings. Positional embeddings are introduced to maintain spatial information, and the transformer encoder then processes this sequence of vectors. 

We use the pre-trained VFMs openly available in \textit{Keras} library for VGG19\footnote{\url{https://keras.io/api/applications/vgg/\#vgg19-function}}, ResNet50\footnote{\url{https://keras.io/api/applications/resnet/\#resnet50-function}}, and EffecientNet\footnote{\url{https://keras.io/api/applications/efficientnet/\#efficientnetb0-function}}. For ViT, we make use of \textit{keras\_vit}\footnote{\url{https://pypi.org/project/keras-vit/}} package and we select ViT\_B32 version. For the VFMs in our study, the number of parameters for each of them are as follows: 143.7M for VGG19, 25.6M for ResNet50, 5.3M for EfficientNet (we select BO version), and 88.22M for ViT.

\begin{figure}[hbt!]    
\centering
      \includegraphics[width=0.30\textwidth, height=0.35\textwidth]{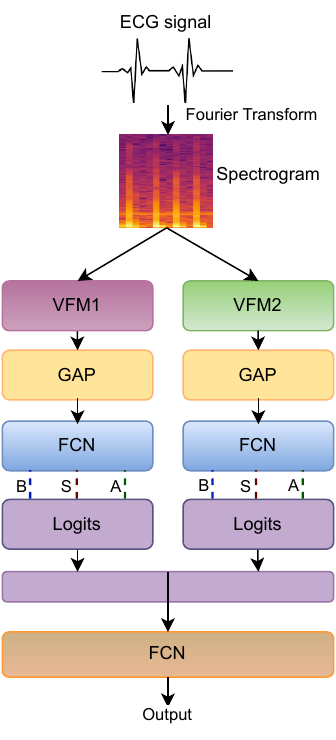}
      \caption{Architecture of \textbf{SONIC}; VFM, GAP, FCN stands for Vision Foundational Model, Global Average Pooling, Fully Connected Network: B, S, A stands for logits of Baseline, Stress, Amusement labels; Here, \textbf{SONIC} is given for 3-class classification, similarly, for 2-class classification is implemented.}
        \label{sonic}     
\end{figure} 
\section{Methodology} 

In this section, we discuss the proposed method, \textbf{SONIC}. With \textbf{SONIC}, we aim to effectively tap into the complementary behavior of the VFMs for improved stress recognition from ECG. The modeling architecture for \textbf{SONIC} is shown in Figure \ref{sonic}. \textbf{SONIC} merges VFMs by fusing their logits, which are the unnormalized predictions generated by each VFM before they are transformed into probabilities through a softmax function. It is important to note that higher logit values indicate a higher likelihood for the corresponding class, portraying greater confidence in the classification. \par

 We anticipate that \textbf{SONIC} can significantly outperform individual VFMs in stress recognition tasks. This expectation stems from its unique capability to capture raw, discriminative information while retaining the distinct contributions of each VFM and harnessing the strengths of individual VFMs while effectively mitigating their weaknesses.

We give a detailed walkthrough of the proposed methodology and its algorithm is shown in Algorithm \ref{training_algorithm}. First, the input ($X$) is sent simultaneously to individual VFM. Each VFM converts the input to a unique feature space tailoring its representative behavior. To be noted, we don't freeze the VFM weights. Each VFM is succeeded by a global pooling (GAP) layer. GAP aids in spatial information retention and parameter reduction by averaging each feature map. Further, we add a fully connected network (FCN) block consisting of two cascading dense layers, the first one with 128 neurons and secondly, 2 (For 2-class classification, representing stress and non-stress labels) or 3 neurons (For 3-class classification, representing baseline, stress and amusement labels). The second layer neurons of the FCN represent the logits for each class. Further, the logits from the VFMs are fused together through concatenation ($Z$). Then passed through FCN consisting of 12 neurons and lastly, the classification head outputs the probabilities corresponding to different classes. We also make use of dropout to reduce overfitting. \par 

\begin{algorithm}
\caption{\textbf{SONIC} Algorithm}\label{training_algorithm}
\begin{algorithmic}[1]
\State \textbf{Input:} Training data $(X, Y)$
\State \textbf{Output:} Trained model

\Procedure{SONIC}{$X, Y, M$}
    \State Send input $X$ to individual VFMs of $M$ and compute logits $Z_1$ and $Z_2$
    \State Concatenate logits: $Z \gets [Z_1, Z_2]$
    \State Initialize FCN parameters: $W_{\text{FC}}, b_{\text{FC}}$
    \For{each training epoch}
        \State $L = - \sum_{i=1}^{N} Y_{i} \log(\hat{Y}_{i})$, here, $N$ stands for the number of training datapoints
        \State Updating VFMs parameters: $W_{\text{VFM}} \gets W_{\text{VFM}} - \alpha \frac{\partial L}{\partial W_{\text{VFM}}}$, $b_{\text{VFM}} \gets b_{\text{VFM}} - \alpha \frac{\partial L}{\partial b_{\text{VFM}}}$
        \State Updating FCN parameters: $W_{\text{FC}} \gets W_{\text{FC}} - \alpha \frac{\partial L}{\partial W_{\text{FC}}}$, $b_{\text{FC}} \gets b_{\text{FC}} - \alpha \frac{\partial L}{\partial b_{\text{FC}}}$, here, $\alpha, W, b$ stands for learning rate, weights, bias 
    \EndFor
    \State \textbf{return} Trained model
\EndProcedure

\end{algorithmic}
\end{algorithm}

\noindent\textbf{Training SONIC:} We train \textbf{SONIC} for 20 epochs for different VFMs combinations with a learning rate as 1e-3 and batch size of 32. We use cross-entropy as the loss function and Adam as the optimizer.

\section{Experiments}

\subsection{Benchmark Datasets:}
We use WESAD (Wearable Stress and Affect Detection) \cite{schmidt2018introducing} dataset for our experiments. It is a widely used benchmark dataset in the field of affective computing and wearable computing. It is often employed to formulate and evaluate algorithms for stress and affect detection using physiological signals collected from wearable sensors. It includes higher resolution physiological data such as ECG, electrodermal activity (EDA), electromyography (EMG), respiratory activity, and temperature. It also includes motion (acceleration) data from chest-worn and wrist-worn devices. The data was collected from 15 participants sampled at 700 Hz. This dataset focuses on both binary class (stress \textit{vs} non-stress) and multi-class classification (amusement \textit{vs} stress \textit{baseline}).

\subsection{Data pre-processing:} 
Here, we give detailed information about the data preprocessing steps. First, raw 1D ECG signals are segmented by creating a window size of 5 seconds and maintaining a window gap of 2 seconds. 
Then we convert them to 2D Spectrograms using Fourier transform\footnote{\url{https://docs.scipy.org/doc/scipy/reference/generated/scipy.signal.spectrogram.html}}. These spectrograms are then reshaped to shape (224, 224). Further, as the VFMs are pre-trained on 3-channel images, we extended the spectrograms to shape (224, 224, 3) by replicating it to three dimensions. The spectrograms are normalized to ensure consistent scaling. We perform 5-fold cross-validation for all the modeling experiments with four-fold as the training set and one-fold as a test set across 5 runs. 
\subsection{Modeling Baselines} 

As baseline experiments for our study, we first fine-tune the VFMs on WESAD for both 2-class and 3-class classification. We fine-tune the VFMs (VGG-19, ResNet50, EfficientNet, ViT) by adding a GAP layer followed by FCN with dense layers of 128 neurons, and 2 or 3 neurons depending on the classification type that represents the classes. We retrain all the weights of the VFMs on WESAD. We employ dropout for mitigating overfitting. We set the number of epochs as 20 and learning rate as 1e-3. We use cross-entropy as the loss function and Adam as the optimizer.

 \subsection{Metrics}

 We make use of accuracy and F1 (macro average) as the metrics for the evaluation of the models.

\begin{table}[hbt!]
\centering
\caption{Performance Evaluation of Models 2-class classification: Scores are average of 5-folds and in \%; \textcolor{red}{A(2)}, \textcolor{red}{F(2)} stands for accuracy and F1 for 2-class classification respectively; \textcolor{blue}{A(3)}, \textcolor{blue}{F(3)} stands for accuracy and F1 for 3-class classification respectively; SOTA(2), SOTA(3) represents the SOTA works on 2-class and 3-class classification respectively}
\label{results}
\begin{tabular}{l|l|l|l|l}   
\hline\hline
\textbf{Model} & \textcolor{red}{\textbf{A(2)}} & \textcolor{red}{\textbf{F(2)}} & \textcolor{blue}{\textbf{A(3)}} & \textcolor{blue}{\textbf{F(3)}}\\\hline 
SOTA (3) \cite{schmidt2018introducing} & 85.44 & 81.31 & 66.29 & 56.03\\
ViT & 87.36 & 84.06 & 72.49 & 54.62\\
SOTA (2)\cite{ishaque2022detecting} & 90.47 & ------ & ------ & ------\\
VGG19 & 90.78 & 88.85 & 73.45 & 54.04\\ 
ResNet50 & 92.39 & 90.80 & 78.63 & 64.93\\
EfficientNet & 92.60 & 91.03 & 80.26 & 72.49\\
SONIC (ViT, VGG-19)  & 92.84 & 91.30 & 73.67 & 61.33 \\
SONIC (ViT, ResNet50)  & 93.69 & 91.87 & 75.39 & 66.86 \\
SONIC (ViT, EfficientNet)  & 93.88 & 91.98 & \cellcolor{green!25}\textbf{85.98} & \cellcolor{green!25}\textbf{79.34} \\
SONIC (VGG-19, ResNet50)  & \cellcolor{green!25}\textbf{94.82} & \cellcolor{green!25}\textbf{93.97} &  83.77 & 73.45 \\
SONIC (VGG-19, EfficientNet)  & \cellcolor{yellow!25}\textbf{97.36} & \cellcolor{yellow!25}\textbf{95.30} & \cellcolor{yellow!25}\textbf{96.76} & \cellcolor{yellow!25}\textbf{95.95}\\
SONIC (ResNet50, EfficientNet) & \cellcolor{blue!25}\textbf{99.36} & \cellcolor{blue!25}\textbf{99.24} &\cellcolor{blue!25}\textbf{97.66} & \cellcolor{blue!25}\textbf{97.10}\\ 

\hline\hline
\end{tabular}
\end{table}

\begin{figure}[t!]
    \centering
     \subfloat[2-Class]{{\includegraphics[width=0.32\textwidth, height = 0.20\textwidth]{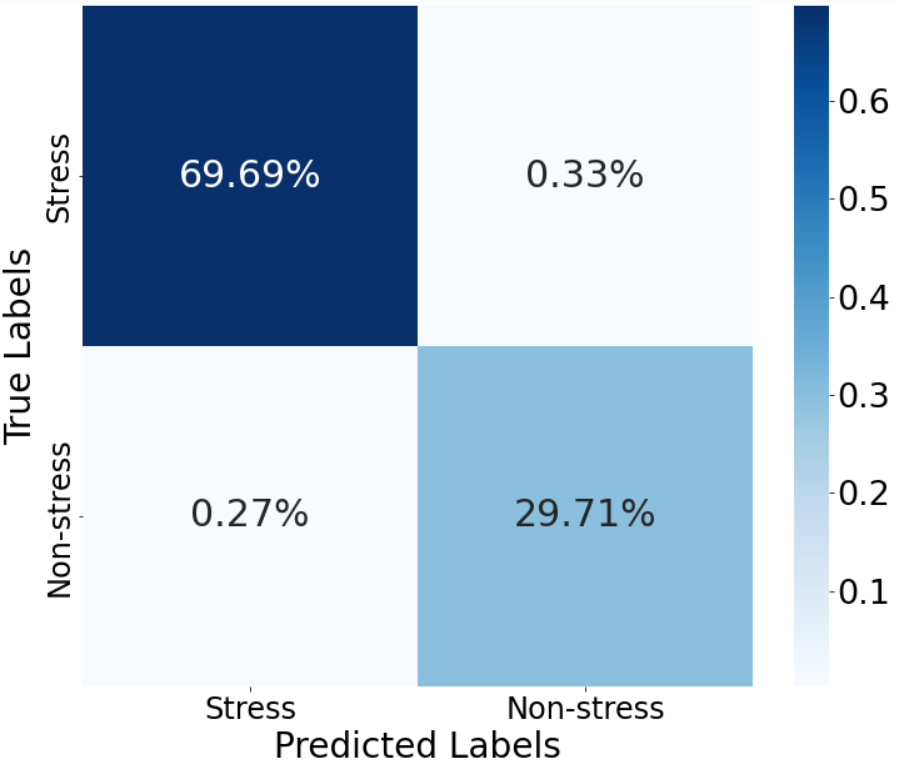}}}\label{fig:uniwcsa}\\
    \subfloat[3-Class]{{\includegraphics[width=0.35\textwidth, height = 0.20\textwidth]{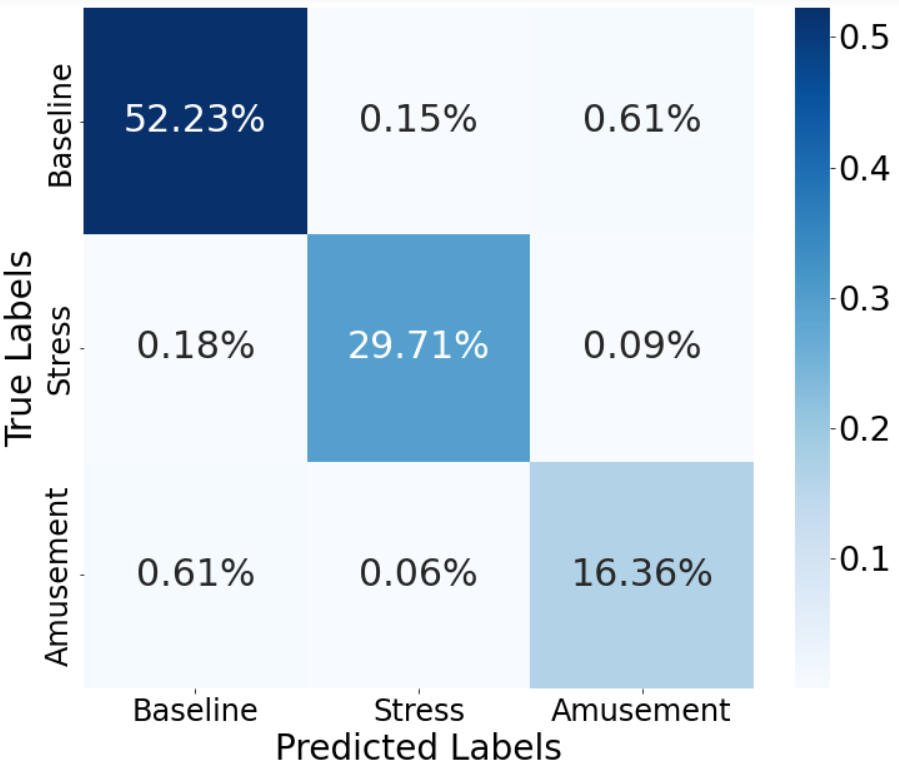}}}\label{fig:unicsa}
   \caption{Confusion Matrix of \textbf{SONIC} with the best VFMs combination (ResNet50, EfficientNet) for 2-class and 3-class classification} 
\label{confmax}
\end{figure}

\begin{figure}[t!]
    \centering
    \subfloat[2-class]{{\includegraphics[width=0.22\textwidth]{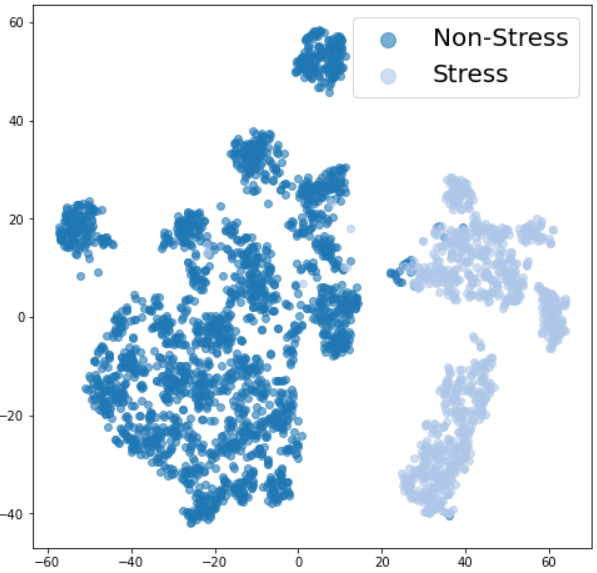}}}\label{fig:unicsa}
    \subfloat[3-class]{{\includegraphics[width=0.22\textwidth]{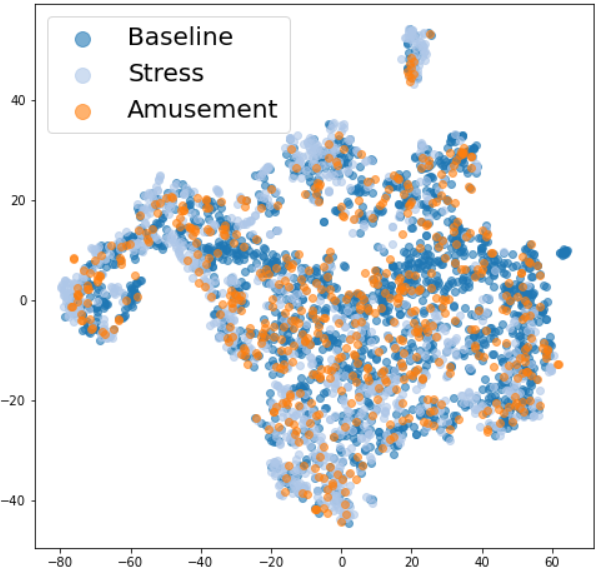}}}\label{fig:uniwcsa}
   \caption{t-SNE plots of EfficientNet for 2-class and 2-Class classification} 
\label{tsne1}
\end{figure}

\begin{figure}[t!]
    \centering
    \subfloat[2-class]{{\includegraphics[width=0.22\textwidth]{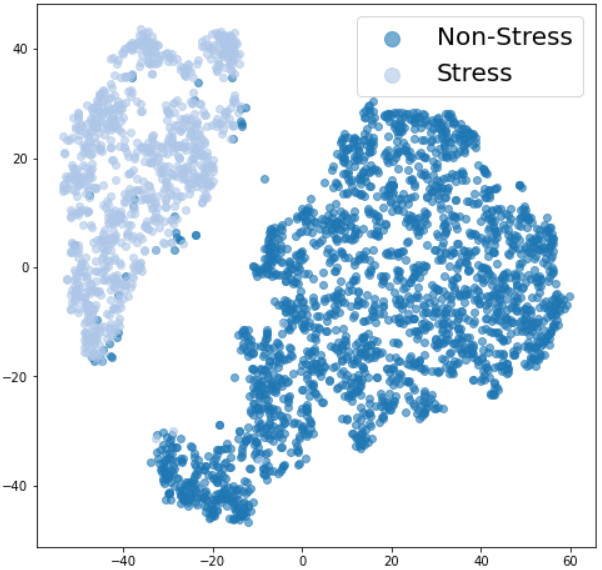}}}\label{fig:unicsa}
    \subfloat[3-class]{{\includegraphics[width=0.22\textwidth]{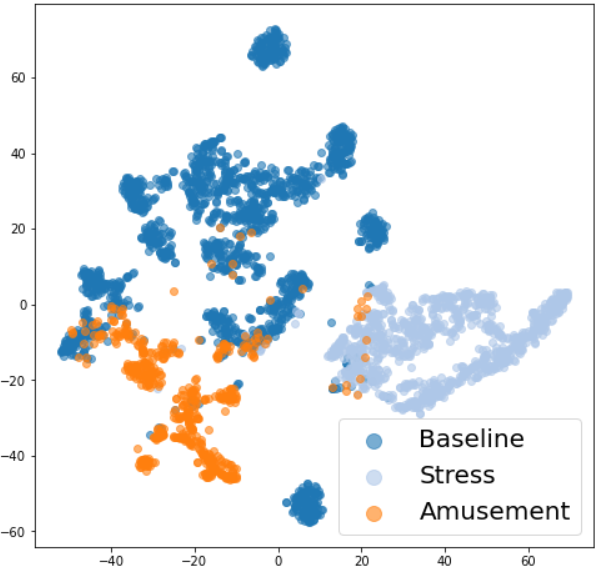}}\label{fig:uniwcsa}}
   \caption{t-SNE plots of \textbf{SONIC} with combination of ResNet and EfficientNet for 2-class and 3-class classfication}
\label{tsne2}
\end{figure}

\subsection{Results}
Table \ref{results} presents the scores obtained from various models. Firstly, we discuss the results obtained by fine-tuning the VFMs individually. We observe that fine-tuning EfficientNet leads to the best performance among the VFMs showing its higher transferability for stress recognition from ECG. This behavior of EfficientNet is also supported by Kornblith et al. \cite{kornblith2019better} which showed that better models in ImageNet transfer better to other downstream tasks. Similar behavior can be observed between ResNet50 and VGG19, as ResNet50 performs better than VGG19 in both 2-class and 3-class classification. \par

We observe that CNN-based VFMs (VGG19, ResNet50, EfficientNet) achieved greater accuracy than transformer-based VFM (ViT), and this can be attributed to CNNs efficacy in capturing local patterns and spatial hierarchies than ViT. In contrast, ViT is designed with self-attention mechanisms which adept at capturing broad-scale relationships within images. \par

From Table \ref{results}, we can see that combining various VFMs leads to improvement in stress recognition performance. This shows that through \textbf{SONIC}, VFMs are showing complementary behavior and improves over their individual performances. Among different combinations, ResNet50 and EfficientNet showed the top most performance, further, proving the effectiveness of \textbf{SONIC}. The confusion matrix of \textbf{SONIC} with the best VFMs combination (ResNet 50 and EfficientNet) for 2-class and 3-class classification is shown in Figure \ref{confmax}. Additionally, we show the t-SNE plots of EfficientNet in Figure \ref{tsne1} and \textbf{SONIC} with ResNet50  and EfficientNet combination in Figure \ref{tsne2} for 2-class and 3-class classification respectively. We observe far better clustering across the classes with \textbf{SONIC}, supporting its topmost performance. For plotting the t-SNE plots, we extract the representations of the trained models for a test set from the penultimate layer before the output layer.

\subsection{Comparison to State-of-the-art}
We compare the results with SOTA works. Table \ref{results} shows the comparison. We obtain around 9\%, 20\% in 2-class and 3-class classification respectively in comparison to previous SOTA models. 

\subsection{Conclusion}
In this work, we introduce \textbf{SONIC}, a novel framework for stress recognition from ECG by combining VFMs through their logits and then training FCN on top of the combined logits. Through extensive experimentation, \textbf{SONIC} demonstrates superior performance compared to individual VFMs on the WESAD benchmark. Remarkably, \textbf{SONIC} achieves SOTA results on WESAD, boasting accuracy rates of 99.36\% and 99.24\% for stress \textit{vs} non-stress classification, and 97.66\% and 97.10\% for amusement \textit{vs} stress \textit{vs} baseline classification, in terms of both accuracy and F1, respectively. We believe our work can open up new possibilities for stress recognition from ECG by harnessing the common complementary feature space of VFMs through the combination of various VFMs. 





\bibliographystyle{IEEEbib}
\bibliography{main}

\begin{thebibliography}{10}

\bibitem{cohen2007psychological}
Sheldon Cohen, Denise Janicki-Deverts, and Gregory~E Miller,
\newblock ``Psychological stress and disease,''
\newblock {\em Jama}, vol. 298, no. 14, pp. 1685--1687, 2007.

\bibitem{hahad2019environmental}
Omar Hahad, J{\"u}rgen~H Prochaska, Andreas Daiber, Thomas Muenzel, et~al.,
\newblock ``Environmental noise-induced effects on stress hormones, oxidative
  stress, and vascular dysfunction: key factors in the relationship between
  cerebrocardiovascular and psychological disorders,''
\newblock {\em Oxidative medicine and cellular longevity}, vol. 2019, 2019.

\bibitem{mravec2020stress}
Boris Mravec, Miroslav Tibensky, and Lubica Horvathova,
\newblock ``Stress and cancer. part i: Mechanisms mediating the effect of
  stressors on cancer,''
\newblock {\em Journal of neuroimmunology}, vol. 346, pp. 577311, 2020.

\bibitem{crosswell2020best}
Alexandra~D Crosswell and Kimberly~G Lockwood,
\newblock ``Best practices for stress measurement: How to measure psychological
  stress in health research,''
\newblock {\em Health psychology open}, vol. 7, no. 2, pp. 2055102920933072,
  2020.

\bibitem{wu23i_interspeech}
Zihan Wu, Neil Scheidwasser-Clow, Karl {El Hajal}, and Milos Cernak,
\newblock ``{Speaker Embeddings as Individuality Proxy for Voice Stress
  Detection},''
\newblock in {\em Proc. INTERSPEECH 2023}, 2023, pp. 1838--1842.

\bibitem{naidu2021stress}
P~Ramesh Naidu, S~Pruthvi Sagar, K~Praveen, K~Kiran, and K~Khalandar,
\newblock ``Stress recognition using facial landmarks and cnn (alexnet),''
\newblock in {\em Journal of Physics: Conference Series}. IOP Publishing, 2021,
  vol. 2089, p. 012039.

\bibitem{tanwar2024attention}
Ritu Tanwar, Orchid~Chetia Phukan, Ghanapriya Singh, Pankaj~Kumar Pal, and
  Sanju Tiwari,
\newblock ``Attention based hybrid deep learning model for wearable based
  stress recognition,''
\newblock {\em Engineering Applications of Artificial Intelligence}, vol. 127,
  pp. 107391, 2024.

\bibitem{schmidt2018introducing}
Philip Schmidt, Attila Reiss, Robert Duerichen, Claus Marberger, and Kristof
  Van~Laerhoven,
\newblock ``Introducing wesad, a multimodal dataset for wearable stress and
  affect detection,''
\newblock in {\em Proceedings of the 20th ACM international conference on
  multimodal interaction}, 2018, pp. 400--408.

\bibitem{chaptoukaev2024stressid}
Hava Chaptoukaev, Valeriya Strizhkova, Michele Panariello, Bianca Dalpaos,
  Aglind Reka, Valeria Manera, Susanne Th{\"u}mmler, Esma Ismailova,
  Massimiliano Todisco, Maria~A Zuluaga, et~al.,
\newblock ``Stressid: a multimodal dataset for stress identification,''
\newblock {\em Advances in Neural Information Processing Systems}, vol. 36,
  2024.

\bibitem{benchekroun2022comparison}
Mouna Benchekroun, Baptiste Chevallier, Hamza Beaouiss, Dan Istrate, Vincent
  Zalc, Mohamad Khalil, and Dominique Lenne,
\newblock ``Comparison of stress detection through ecg and ppg signals using a
  random forest-based algorithm,''
\newblock in {\em 2022 44th Annual International Conference of the IEEE
  Engineering in Medicine \& Biology Society (EMBC)}. IEEE, 2022, pp.
  3150--3153.

\bibitem{giannakakis2019novel}
Giorgos Giannakakis, Eleftherios Trivizakis, Manolis Tsiknakis, and Kostas
  Marias,
\newblock ``A novel multi-kernel 1d convolutional neural network for stress
  recognition from ecg,''
\newblock in {\em 2019 8th International Conference on Affective Computing and
  Intelligent Interaction Workshops and Demos (ACIIW)}. IEEE, 2019, pp. 1--4.

\bibitem{zhang2021real}
Pengfei Zhang, Fenghua Li, Rongjian Zhao, Ruishi Zhou, Lidong Du, Zhan Zhao,
  Xianxiang Chen, and Zhen Fang,
\newblock ``Real-time psychological stress detection according to ecg using
  deep learning,''
\newblock {\em Applied Sciences}, vol. 11, no. 9, pp. 3838, 2021.

\bibitem{zhang2021psychological}
Pengfei Zhang, Fenghua Li, Lidong Du, Rongjian Zhao, Xianxiang Chen, Ting Yang,
  and Zhen Fang,
\newblock ``Psychological stress detection according to ecg using a deep
  learning model with attention mechanism,''
\newblock {\em Applied Sciences}, vol. 11, no. 6, pp. 2848, 2021.

\bibitem{seo2019deep}
Wonju Seo, Namho Kim, Sehyeon Kim, Chanhee Lee, and Sung-Min Park,
\newblock ``Deep ecg-respiration network (deeper net) for recognizing mental
  stress,''
\newblock {\em Sensors}, vol. 19, no. 13, pp. 3021, 2019.

\bibitem{behinaein2021transformer}
Behnam Behinaein, Anubhav Bhatti, Dirk Rodenburg, Paul Hungler, and Ali Etemad,
\newblock ``A transformer architecture for stress detection from ecg,''
\newblock in {\em Proceedings of the 2021 ACM International Symposium on
  Wearable Computers}, 2021, pp. 132--134.

\bibitem{ishaque2022detecting}
Syem Ishaque, Naimul Khan, and Sri Krishnan,
\newblock ``Detecting stress through 2d ecg images using pretrained models,
  transfer learning and model compression techniques,''
\newblock {\em Machine Learning with Applications}, vol. 10, pp. 100395, 2022.

\bibitem{simonyan2014very}
Karen Simonyan and Andrew Zisserman,
\newblock ``Very deep convolutional networks for large-scale image
  recognition,''
\newblock {\em arXiv preprint arXiv:1409.1556}, 2014.

\bibitem{dosovitskiy2020image}
Alexey Dosovitskiy, Lucas Beyer, Alexander Kolesnikov, Dirk Weissenborn,
  Xiaohua Zhai, Thomas Unterthiner, Mostafa Dehghani, Matthias Minderer, Georg
  Heigold, Sylvain Gelly, et~al.,
\newblock ``An image is worth 16x16 words: Transformers for image recognition
  at scale,''
\newblock {\em arXiv preprint arXiv:2010.11929}, 2020.

\bibitem{bakasa2023stacked}
Wilson Bakasa and Serestina Viriri,
\newblock ``Stacked ensemble deep learning for pancreas cancer classification
  using extreme gradient boosting,''
\newblock {\em Frontiers in Artificial Intelligence}, vol. 6, 2023.

\bibitem{alotaibi2023vit}
Amira Alotaibi, Tarik Alafif, Faris Alkhilaiwi, Yasser Alatawi, Hassan
  Althobaiti, Abdulmajeed Alrefaei, Yousef Hawsawi, and Tin Nguyen,
\newblock ``Vit-deit: An ensemble model for breast cancer histopathological
  images classification,''
\newblock in {\em 2023 1st International Conference on Advanced Innovations in
  Smart Cities (ICAISC)}. IEEE, 2023, pp. 1--6.

\bibitem{he2016deep}
Kaiming He, Xiangyu Zhang, Shaoqing Ren, and Jian Sun,
\newblock ``Deep residual learning for image recognition,''
\newblock in {\em Proceedings of the IEEE conference on computer vision and
  pattern recognition}, 2016, pp. 770--778.

\bibitem{tan2019efficientnet}
Mingxing Tan and Quoc Le,
\newblock ``Efficientnet: Rethinking model scaling for convolutional neural
  networks,''
\newblock in {\em International conference on machine learning}. PMLR, 2019,
  pp. 6105--6114.

\bibitem{vaswani2017attention}
Ashish Vaswani, Noam Shazeer, Niki Parmar, Jakob Uszkoreit, Llion Jones,
  Aidan~N Gomez, {\L}ukasz Kaiser, and Illia Polosukhin,
\newblock ``Attention is all you need,''
\newblock {\em Advances in neural information processing systems}, vol. 30,
  2017.

\bibitem{kornblith2019better}
Simon Kornblith, Jonathon Shlens, and Quoc~V Le,
\newblock ``Do better imagenet models transfer better?,''
\newblock in {\em Proceedings of the IEEE/CVF conference on computer vision and
  pattern recognition}, 2019, pp. 2661--2671.

\end{thebibliography}

\end{document}